\documentclass[12pt]{article}
\usepackage[tbtags]{amsmath}
\usepackage{graphicx}
\usepackage{amssymb}
\usepackage{epstopdf}
\usepackage{cite}
\usepackage{topcapt}
\usepackage{booktabs}

\newcommand{\be}{\begin{equation}}
\newcommand{\la}{\langle}
\newcommand{\ra}{\rangle}
\newcommand{\ee}{\end{equation}}
\title{From Entropy to Information: Biased Typewriters and the Origin of Life}
\author{Christoph Adami$^{1,2,3}$ and Thomas LaBar$^{1,3}$}

\begin{document}
\date{}
\maketitle

\begin{flushleft}
$^{\bf{1}}$ Department of Microbiology and Molecular Genetics, Michigan State University, East Lansing, Michigan, USA.
\\
$^{\bf{2}}$ Department of Physics and Astronomy, Michigan State University, East Lansing, Michigan, USA.\\
$^{\bf{3}}$ BEACON Center for the Study of Evolution in Action, Michigan State University, East Lansing, Michigan, USA.\\

$^{\ast}$ E-mail: adami@msu.edu
\end{flushleft}


{\bf Introduction} So much has been written about the possible origins of life on Earth (see, e.g., the popular books~\cite{Morowitz2004,Deamer1994,DeDuve1995,Koonin2011}) that it sometimes seems that---barring an extraordinary breakthrough in experimental biochemistry (for example~\cite{Pateletal2015}), or else the discovery of the remnants of an ancient biochemistry~\cite{Daviesetal2009}---nothing new can be said about the problem. But such a point of view does not take into account that perhaps not all the tools of scientific inquiry have been fully utilized in this endeavor to unravel our ultimate origin on this planet. Indeed, Origin-of-Life research has historically been confined to a fairly narrow range of disciplines, such as biochemistry and  geochemistry. Today, a much broader set of tools is being unleashed on this problem, including mathematical~\cite{Vetsigianetal2006,Smith2008,England2013} and computational approaches~\cite{Segreetal2000,NowakOtsuki2008,Vasasetal2012,Walkeretal2012,Mathisetal2015}. Computational approaches to the study of possible origins of life are often derided because they lack a particular feature of biochemistry,  or ``because they do not take into account the specific properties of individual organic compounds and polymers"~\cite{LazcanoMiller1996}. Such a point of view ignores the possibility that life may not a feature that is dependent on a particular biochemistry~\cite{Benneretal2004}, but could instead be a feature of {\em any} chemistry that is capable of encoding information. 

If the one invariant in life is information (information about how to replicate, that is), it then becomes imperative to understand the general principles by which information could arise by chance. It is generally understood that evolution, viewed as a computational process~\cite{Adami1998,Mayfield2013} leads to an increase in information on average. The amount of information that evolution has accumulated to date differs from organism to organism of course, and precise numbers are not known. A rough estimate of the amount of information stored in an organism's genome can be obtained by calculating the amount of functional DNA in an organism\footnote{It is not necessary to consider epigenetic variation in the estimate of information content, as all epigenetic changes are performed by enzymes whose information is already stored within DNA.}. The general idea here is that only functional DNA can be under selection, as after all information is that which guarantees survival~\cite{Adami2002b,Adami2012}. 
For humans (assuming a functional percentage of about 8\%~\cite{Randsetal2014}), this means that our DNA codes for about half a billion bits\footnote{This number is (given the functional percentage of 8\%) an upper limit on the information content, as protein coding regions display considerable variation and redundancy, which lowers information. However, as open reading frames only account for 1\% of the human genome and regulatory sequences (the other 7\%) are much less redundant, the true information content of human DNA is likely not much lower.}.

Almost all of the information contained in our genome (and any other organism's) owes its existence to the evolutionary process. But the algorithm that is evolution cannot be at work in the absence of replication, and therefore cannot explain the origin of life. It is in principle possible that the first replicator did not originate on Earth but rather arrived on Earth from extra-terrestrial sources~\cite{Arrhenius1908,HoyleWickramasinghe1981,Wickramasinghe2011}. Even if that was the case, such an origin story does not obviate the need for emergence {\em somewhere}, so we may ask generally: ``What is the likelihood of spontaneous emergence of information?".  The question in itself is not new, of course. Howard Pattee asked as early as 1961, shortly after the discovery of the structure of DNA (but before the discovery of the genetic code)~\cite{Pattee1961}:
\begin{quote} 
(1) How did a disordered collection of elements which forms sequences with no restrictions produce, with reasonable probability, enough initial order to result in the general property of self-replication? (2) Assuming the general property of self-replication for all sequences, how did those particular sequences which now exist arise, with reasonable probability, from the set of all possible sequences?
\end{quote}

In order to estimate the likelihood of spontaneous emergence of a self-replicator, it is necessary to estimate the {\em minimal information} necessary to replicate, because the length of the sequence is not a good indicator of fitness.  A quick gedankenexperiment can clarify this. Imagine that a symbolic sequence (written using ASCII characters) can replicate if and only if anywhere on the string the exact sequence {\tt origins} appears. This is a 7 letter sequence, and the total number of possible sequences of length 7 is $26^7$, or about 8 billion. The likelihood to find this sequence by chance if a billion sequences are tried is, obviously, about 1 in 8. But suppose we try sequences of length 1,000. If we only ask that the word appears {\em anywhere} in the sequence, increasing sequence length obviously increases both the number of possible sequences and the number of self-replicators. Thus, the likelihood to find a self-replicator does not scale exponentially with the length of the sequence (it does not become $26^{-1,000}$), but rather with the {\em information content} of the sequence (as we will see momentarily). In the present example, the information content is clearly 7 letters. But how do you measure the information content of biomolecules?

{\bf Information content of biomolecules} Generally speaking, the information content of a symbolic sequence is equal to the amount of uncertainty (about a particular ensemble) it can reduce. This information can be written mathematically in terms of the entropy of the ensemble (described by the random variable $X$ that can take on states $x_1,..,x_n$ with probabilities $p_1,...,p_n$
\be
H(X)=-\sum_{i=1}^n p_i\log p_i
\ee
and the conditional entropy $H(X|s)$, where $s$ is the sequence whose information content we would like to measure, as
\be
I(s)=H(X)-H(X|s)\;.
\ee
The latter entropy is given by the conditional entropy distribution $p_{i|s}$ instead. So, for example, the sequence {\tt Colonel Mustard} reduces the uncertainty about the identity of the murderer in a popular board game from $\log_26\approx 2.83$ bits to zero (as there are a priori six suspects, and the sequence fingers the perpetrator), so the information content is 2.83 bits. The sequence length, on the contrary, is 15 (counting the space as a symbol), which translates to $15 \log_2(27)\approx 71.3$ bits. Thus, sequence length and information content can be very different: information is about something, while sequence length is just entropy.  

Unfortunately, we cannot measure the information content of biomolecules in the same manner, because we do not know the entropy of the ensemble that the biomolecular sequence is information about. Let us call this random variable $E$ (for ``environment"), as it represents the environment within which the sequence is functional, in the same sense that $X$ above was the environment within which the sequence {\tt Colonel Mustard} is functional. 
However, an information-theoretical ``trick" allows us to make progress. Let $s$ be a functional biomolecule (a polymer of length $L$), and its information content (per the formula above)
\be
I(s)=H(E)-H(E|s)\;, \label{inf}
\ee
that is, it is the entropy of the ``world" minus the entropy of the world given that we know $s$. We can also define the average information content as
\be
\la I\ra =\sum_sp(s)I(s)=H(E)-H(E|S)=H(E:S)
\;,
\ee
where $H(E:S)$ is the shared entropy between environment and sequences,  but again that formula is not useful because we do not know $H(E)$. However, the formula can also be written as
\be
\la I\ra=H(S)-H(S|E)
\ee
in terms of the entropy of sequences $H(S)$ and the conditional entropy of the sequences given an average environment. This is also not useful, as the world is not an average of environments, but one very particular one $E=e$. Could we write this in terms of a difference of entropies as in (\ref{inf})? We then would guess that
\be
I(s)=H(S)-H(S|e)\;, \label{inf1}
\ee
but equation (\ref{inf1}) is not mathematically identical to (\ref{inf}), as the identity only holds for the averages. However, Eq.~(\ref{inf1}) can be derived from an approach embedded in Kolmogorov complexity theory~\cite{Adami1998,AdamiCerf2000,Adami2002b}, where that equation represents the ``physical complexity" of the sequence. Furthermore, (\ref{inf1}) is practical to the extent that its value can be estimated. For example, as $S$ is the ensemble of sequences, its entropy is simply given by $\log N$, where $N$ is the total number of sequences of that length (it is possible to extend this formalism to sequences of varying length). Sequences with an arbitrary function in environment $E=e$ have an entropy  smaller than $\log N$. Let us imagine that the number of polymers with that function (in $e\in E$) is $N_e$ (with $N_e\ll N$). Then (here we specify the base of the logarithm by the number of possible monomers $D$)
\be
I(s)=-\log_D \frac{N_e}N \label{infofrac}
\ee
which, it turns out, is identical to Szostak's ``functional complexity" measure~\cite{Szostak2003}. It allows us to quantify the information content of a biomolecular sequence if the ``density" of functional sequences $N_e/N$ is known, and makes it possible to calculate the likelihood of emergence (by chance), of a molecule with information content $I$. As the likelihood must be given by the density of molecules of that type within the set of all molecules of that length, 
we find
\be
P=\frac{N_e}N=D^{-I}\;, \label{exp}
\ee
where the relationship to information content follows directly from (\ref{infofrac}). Thus we see (as advertised earlier), that this likelihood {\em only} depends on the information content of the sequence, but not on its length. Below, we will test this prediction using the digital life system Avida and find it violated. However, the origin of this apparent violation is easily tracked down, and we are confident that the equality holds exactly in principle.
 
{\bf Testing the likelihood of emergence by chance}. We first tested the likelihood to find the sequence {\tt origins} by creating random ASCII polymers of length 7 using an alphabet of $D=26$ (no spaces or other punctuation), and where each symbol was drawn from a uniform distribution over the letters a-z. When testing a billion sequences we did not find {\tt origins}, which is in accord with the probability $P=26^{-7}$ calculated above. Note that for ASCII strings (unlike the biomolecules) there is never any redundancy, so that $N_e=1$ always. We then randomly searched for self-replicating sequences within the digital chemistry of the Avida Artificial Life system. \cite{AdamiBrown1994,Adami1998,OfriaWilke2004,Ofriaetal2009}.
In Avida, ASCII sequences {\em can self-replicate}, but only because these sequences are translated to instructions that are executed on virtual CPUs. In this sense, the sequences are really self-replicating computer programs, and because these sequences can mutate as they are copied, they evolve in a strictly Darwinian manner (see Table 1 for the arbitrary assignment of ASCII letters to avidian instructions). 
\begin{table}[htbp]
   \centering
   \begin{tabular}{@{} lll @{}} 
      \toprule
      Instruction    & Description & Symbol\\
      \midrule
nop-A    & no operation (type A) & a \\
nop-B   & no operation (type B) & b \\
nop-C   & no operation (type C) & c \\
if-n-equ & Execute next instruction only-if ?BX? does not equal complement & d\\
if-less & Execute next instruction only if ?BX? is less than its complement & e\\
if-label & Execute next instruction only if template complement was just copied & f\\
mov-head & Move instruction pointer to same position as flow-head & g\\
jmp-head & Move instruction pointer by fixed amount found in register CX & h\\
get-head & Write position of instruction pointer into register CX & i\\
set-flow & Move the flow-head to the memory position specified by ?CX? & j\\
shift-r & Shift all the bits in ?BX? one to the right & k\\
shift-l & Shift all the bits in ?BX? one to the left & l\\
inc & Increment ?BX? & m\\
dec & Decrement ?BX? & n\\
push &       Copy value of ?BX? onto top of  current stack & o\\
pop & Remove number from current stack and place in ?BX? & p\\
swap-stk & Toggle the active stack & q\\
swap & Swap the contents of ?BX? with its complement & r\\
add & Calculate  sum of BX and CX; put  result in ?BX? & s\\
sub & Calculate  BX minus CX; put result in ?BX? & t\\
nand & Perform bitwise NAND on BX and CX; put  result in ?BX? & u\\
h-copy &  Copy instruction from read-head to write-head and advance both & v\\
h-alloc & Allocate memory for offspring & w\\
h-divide & Divide off an offspring located between read-head and write-head & x \\
IO &  Output value ?BX? and replace with new input & y\\
h-search & Find complement template and place flow-head after it & z\\
      \bottomrule
   \end{tabular}
   \caption{Instruction set of the avidian programming language used in this study. The notation ?BX? implies that the command operates on a register specified by the subsequent nop instruction (for example, nop-A specifies the AX register, and so forth). If no nop instruction follows, use the register BX as a default. More details about this instruction set can be found in~\cite{Ofriaetal2009}.}
   \label{avidainst}
\end{table}
The Avida system has been used for over 20 years to test evolutionary dynamics (see, for example, the review~\cite{Adami2006} covering mostly the first ten years), and the likelihood of emergence of functional information (but not self-replication) has been studied in this system before~\cite{Hazenetal2007}. (See also~\cite{Pargellis2003} for an investigation of spontaneous emergence of digital life in a related digital system).  
\begin{figure}[htbp] 
   \centering
   \includegraphics[width=3.5in]{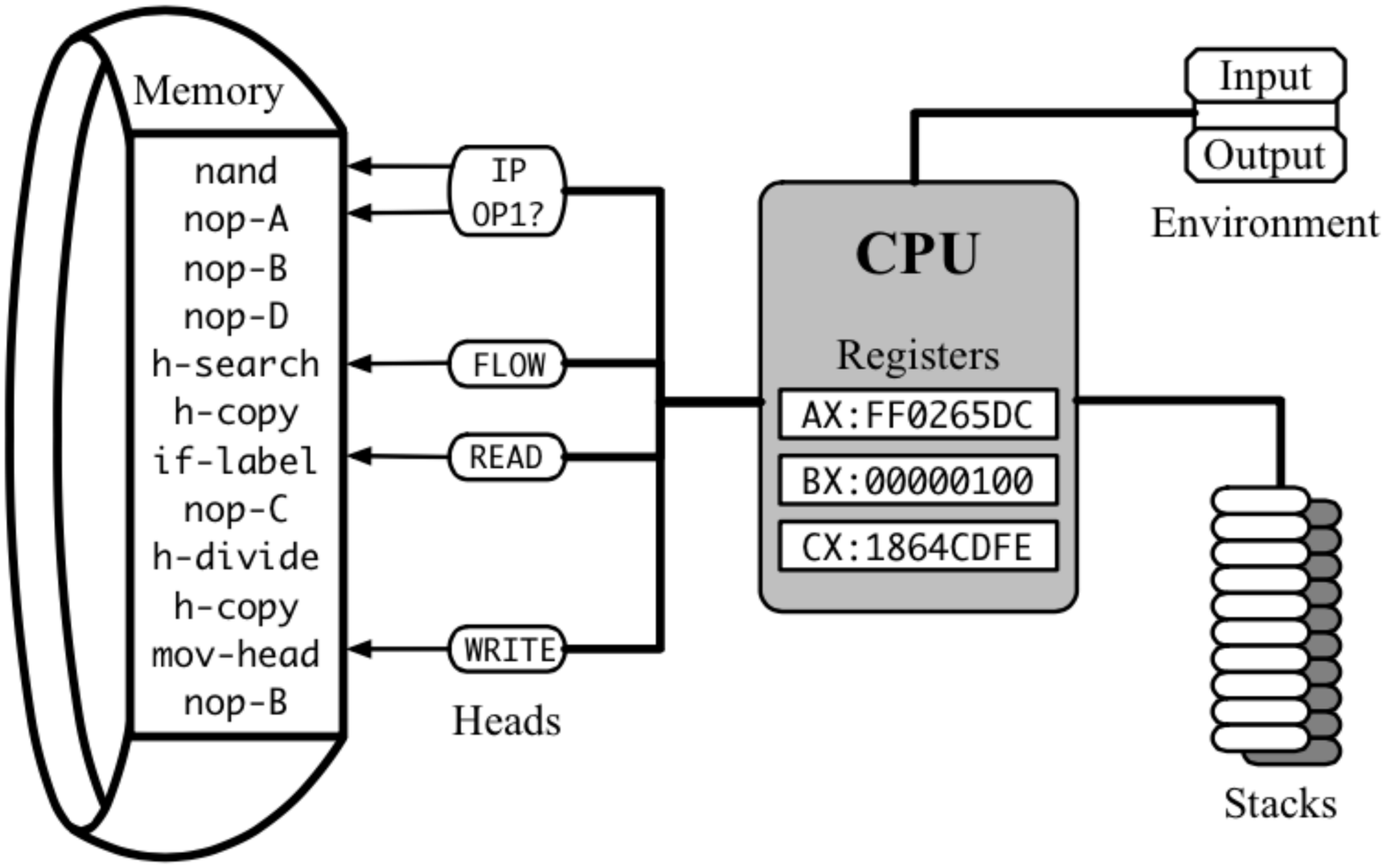} 
   \caption{Sketch of the avidian CPU, executing a segment of code. The CPU uses three registers (AX,BX,CX) and an instruction pointer (IP) that reads the program into the CPU.  A read-head, a write-head, and a flow-head are used to specify positions in the CPU's memory. For example, the `copy' command reads from the read-head and writes to the write-head, while `jump'-type statements move the instruction pointer to the flow-head. The CPU uses two stacks to simulate an ``infinite Turing tape", and input/output buffers to communicate with its environment (reproduced from~\cite{Ofriaetal2009}, with permission). }
   \label{avida}
\end{figure}

The likelihood that any particular sequence coded within 26 instructions can replicate depends strongly on the meaning of each instruction. If a single letter (monomer) were to be interpreted as ``replicate the entire sequence it is in", then self-replicators would be very easy to find. Over the years of development of Avida, the meaning of each symbol has changed as the instruction set itself has changed over time, so the absolute values for the information content of self-replicators may also change in the future. We are here only interested in the rate at which self-replicators can be found in relationship to the information content, and how this rate depends on other factors in the environment that can be modified. Translated to a search for the origins of life, we are interested in how local (environmental) conditions can favorably increase the likelihood to find a self-replicator with information content $I$ purely by chance.

We first focused on avidian sequences constrained to length $L=15$, as there already is a hand-written standard replicator of that length in Avida, given by the string {\tt wzcagczvfcaxgab}. If every instruction in this replicator were information, the likelihood of finding it by chance would be $26^{-15}\approx 6\times 10^{-22}$. Even if we tested a million sequences per second per CPU (central processing unit), on 1,000 CPUs running in parallel, we only would expect to find a single self-replicator in about 50,000 years of continuous search.  We tested one billion sequences of $L=15$ and found 58 self-replicators (all of them unique) by chance, indicating that the information content of self-replicators is vastly smaller than 15 mers. Indeed, we can estimate the information content as
\be
I(15)=-\log_D (58\times 10^{-9})\approx 5.11\pm 0.04\ {\rm mers}\;,
\ee
with a one-$\sigma$ error. Here, the `mer' is a unit of information obtained by taking logarithms to the base of the alphabet size, so that a single monomer has up to one mer of entropy~\cite{Adami2002b,Adami2012}. This means that, within the replicating 15-mers, only about 5 of those 15 mers are information. 

We next tested the information content of sequences constrained to several different lengths. Among a billion random sequences of $L=30$, we found 106 replicators, which translates to
\be
I(30)=-\log_D (106\times 10^{-9})\approx 4.93\pm 0.03\ {\rm mers}\;,
\ee
which is significantly different from $I(15)$. In fact, the calculated information content suggests that perhaps replicators of length five or six might exist, but an exhaustive search of all 11,881,376 $L=5$ sequences and all 308,915,776 $L=6$ sequences reveals this not to be the case. When searching a billion sequences of $L=8$ we found 6 unique self-replicators, implying an information content
\be
I(8)=-\log_D (6\times 10^{-9})\approx 5.81\pm 0.13\ {\rm mers}\;.
\ee
\begin{figure}[!b] 
   \centering
   \includegraphics[width=4in]{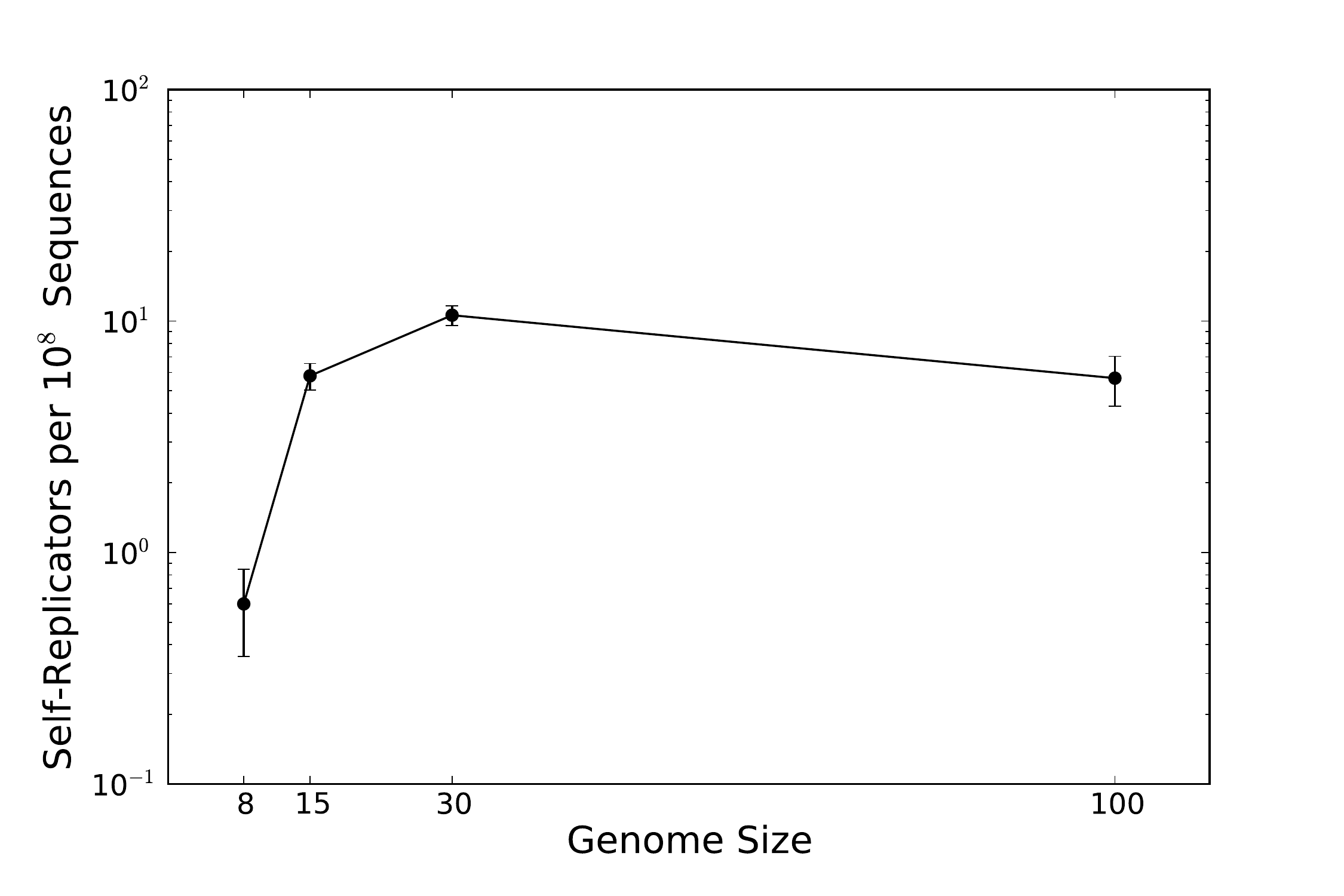} 
   \caption{Number of self-replicators per $10^8$ found for various genome sizes using an unbiased (uniform) probability distribution of monomers. The number of self-replicators per $10^8$ for $L=100$ is estimated from sampling 300 million sequences only (all others used samples of $10^9$).  Error bars are standard deviations.}
   \label{reps}
\end{figure}
The six sequences we found are  {\tt qxrchcwv, vxfgwjgb, wxvxfggb, vhfgxwgb, wxrchcvz}, and {\tt wvfgjxgb}.

We can understand this trend of decreasing information content with increasing length (violating Eq.~(\ref{exp})) as a consequence of the way we treat avidian sequences, namely as having a beginning and an end. Indeed, while the genome itself is circular, execution always begins at a marked instruction. We can see this effect at work using the example {\tt origins} sequence that we used before. If we add a single letter to the 7-mer {\tt origins}, the number of sequences that spell the word increases by 52 (adding the letter to the beginning or the end of the word), while the total number of possible sequences only increases by 26. Thus, the density of self-replicators increases with length, leading to a decrease of information. 

We tested whether this decrease of information with increasing sequence length would continue,  by testing 300 million sequences of length 100. We found 17 self-replicators among this set, which translates to $I(100)=5.10\pm0.09$ mers and suggests that not only does the trend not continue (which of course would have been absurd), but may reverse itself. There is a subtle information-theoretic reason for an increase in information with increasing sequence length. Suppose that there is a single instruction that could abrogate self-replication if it is to be found anywhere within the sequence, when in its absence the sequence replicates (a `kill' instruction, so to speak). Even though such an instruction is obviously not information about how to self-replicate, its needed absence actually {\em is} information. When the sequence length increases, the presence of such a `kill' instruction becomes more and more likely, and therefore the absence of the instruction over the increasing sequence length represents an increase in information. This is the trend suggested in Fig.~\ref{reps}.

{\bf Biased typewriters} In a sense, the random search for self-replicators is very inefficient: it is known that functional molecular sequences cluster in genetic space, while vast regions of that space are devoid of function. Yet, the random generation of sequences searches all of genetic space evenly. Is there a way to focus the search more on sequences that are likely to be functional? It turns out there is, and this method only requires the generation of monomers using a biased probability density function that more resembles that generated by functional sequences~\cite{Adami2014}. We first present a simple example (the biased typewriter), and then outline the theory behind the enhanced search.

\begin{figure}[htbp] 
   \centering
   \includegraphics[width=4in]{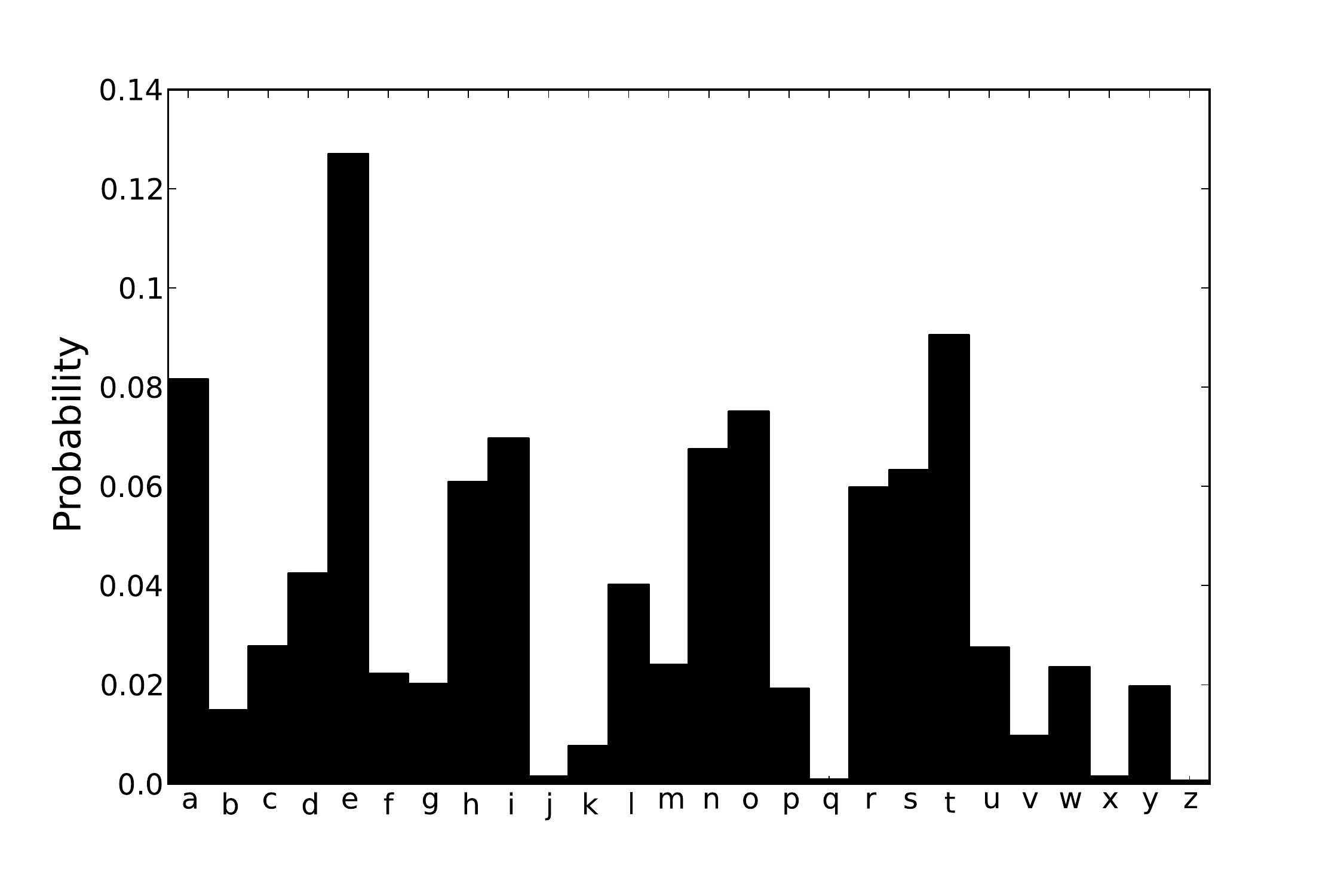} 
   \caption{The probability distribution of letters in the English language. Data from \cite{Lewand2000}.}
   \label{english}
\end{figure}

Words in the English language have a very characteristic letter-frequency distribution that makes it possible to distinguish English text from random sequences of letters, and even text written in different languages. Fig.~\ref{english} (using data from~\cite{Lewand2000})
shows the frequency distribution of letters in English text, showing that  `e' appears more frequently than `t', which itself is more frequent than `a' and so on. 
As this is the expected frequency of letters in English, a focused search should generate words with these expected frequencies  that is, the `monomers' of English words should be generated with the frequency distribution in Fig.~(\ref{english}), rather than uniformly. When we did this for 1 billion sequences of seven letters, we found {\tt origins} twice. How large is the expected increase in likelihood?

We saw earlier that the information content of sequence $s$ can be written as
\be
I(s)=-\log \frac{N_e}N \;,
\ee
which itself is an approximation of the form
\be
I(s)=H(S)-H(S|e)\;, \label{info}
\ee
assuming that the distribution of functional sequences in genetic space is uniform\footnote{The distinction between the entropy written as $\log N_e$ or else as $-\sum_sp(s|e)\log p(s|e)$ can viewed as the same distinction that is made in thermodynamics, where the former is known as the entropy in the ``micro-canonical ensemble", whereas the latter entropy pertains to a ``canonical ensemble" if $p(s|e)$ is the canonical distribution, see, e.g.~\cite{Reif1965}.}. The remaining entropy (given the current environment $E=e$) $H(S|e)$ is not known a priori, but we can estimate it. This entropy of the polymer $s\in S$ can be written in terms of the entropy of monomers, the shared entropy of all monomer pairs, triplets, and so on, using a formula that was first derived by Fano in a very different context~\cite[p. 58]{Fano1961}:
\be
H=\sum_{i=1}^L H(i)-\sum_{i>j}^LH(i:j)+\sum_{i>j>k}^LH(i:j:k)-\cdots  \label{full}
\ee
where $H(i)$ is the entropy of the $i$th monomer, $H(i:j)$ is the shared entropy between the $i$th and $j$th monomer, and so on.  
The sum in (\ref{full}) has alternating signs of correlation entropies, culminating with a term $(-1)^{L-1}H(1:2:3:\cdots :L)$.  
The per-site entropies $H(i)$ can easily be obtained if ensembles of functional molecular sequences are known, as multiple alignment  of these sequences can give us the probability distribution $p(i)$ at each site. The pairwise entropies $H(i:j)$ are important too, in particular if the monomers in the polymer interact functionally, as is often the case if the sequence folds into a structure~\cite{GuptaAdami2014}. Here we will use only the first term in (\ref{full}) to discuss the likelihood of information emergence by chance, but we will discuss the effect of neglecting the other terms below.

In the following, we will use the symbol $I_0$ for the information content of a self-replicator measured using only the first term in (\ref{full}), given by
\be
I_0=L-\sum_{i=1}^L H(i)\;. \label{inf0}
\ee
The first term in (\ref{inf0}) is, of course, the first term in (\ref{info}) if $H(S)=\log(N)$ and we agree to take logarithms to the base of the size of the alphabet. In that case, $\log_D N=\log_D D^L=L$. Using this expression, the likelihood to find self-replicators by chance is approximated as 
\be
P_0=D^{-I_0}=D^{-L+\sum_{i=1}^L H(i)}\;. \label{p0}
\ee
Let us define the ``average biotic entropy" $H_b$ as the average entropy-per-site for functional sequences (hence the name ``biotic")
\be
H_b=\frac1L\sum_i^LH(i)
\ee
We distinguish this biotic entropy from the ``abiotic" entropy $H_\star$, which is the entropy per-site within a sequence assembled at random. If each monomer appears with uniform probability, then the abiotic entropy is maximal: $H_\star=1$. Using this definition, we can write (\ref{p0}) as
\be
P_0=D^{-L(1-H_b)}\;.
\ee
If we were to generate ASCII sequences with a probability distribution obtained from English words (the equivalent of the biotic sample, see Fig.\ref{english}), the abiotic entropy would be smaller than 1 (namely $H_\star\approx 0.89$, the entropy of the distribution in Fig.~\ref{english}) while the biotic entropy must be zero, as there is only a single {\tt origins} among 7-mers. Using the probability distribution of letters in English rather than the uniform distribution raises the probability to find the 7-mer {\tt origins} to
\be
P_\star =26^{-7(0.89)}\;.
\ee
This seems like a small change, but the mean number of successes out of $10^9$ tries is increased from about 1 in 8 billion to 1.53 per billion. And indeed, we found the word twice when searching a billion sequences with the biased distribution shown in Fig.~\ref{english}. Note, however, that the entropy of English is equal to the entropy $\frac1L\sum_i^LH(i)$ only if sequences cannot be aligned, and therefore that all $H(i)\approx H_\star$. 

\begin{figure}[htbp] 
   \centering
   \includegraphics[width=4in]{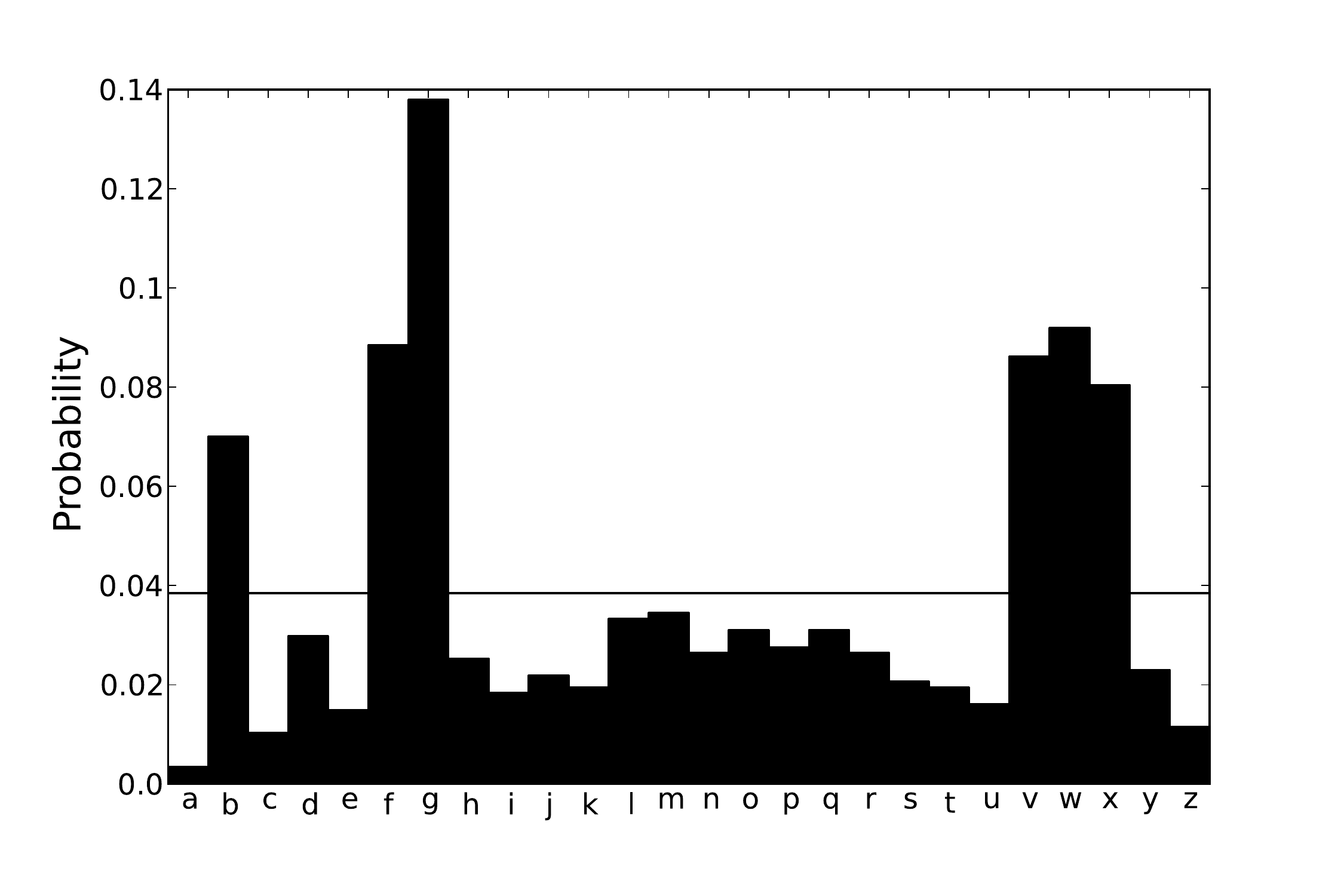} 
   \caption{The biased probability distribution $p_\star$ of Avida instructions obtained from the genomes of 58 randomly generated $L=15$ replicators (the meaning of each letter is described in Table~\ref{avidainst}). The solid black line represents the probability for a uniform distribution.}
   \label{avida15}
\end{figure}

Can searching with a biased probability distribution increase the chance of finding a self-replicator in Avida? We first took the 58 self-replicators we found when searching $L=15$ sequences, and created a monomer-probability distribution $p_\star$ out of them. This distribution in Fig.~\ref{avida15} shows that within these randomly created replicating sequences, the 26 instructions appear far from uniformly in the sequence (as of course is expected), in the same way as English (because it conveys information) has a non-uniform letter distribution.  
The entropy of the distribution shown in Fig.~\ref{avida15} is $H_\star=0.91$ mers. According to the approximation we made above, biasing the monomer creation process using this particular probability distribution should lead to an enhancement $E$ of the likelihood of finding a self-replicator
\be
E=\frac{P_\star}{P_0}\approx\frac{D^{-L(H_\star-H_b)}}{D^{-L(1-H_b)}}=D^{L(1-H_\star)}\;. \label{enh}
\ee
Eq.~(\ref{enh}) suggests that the enhancement factor $E$ only depends on the bias in the distribution and the length of sequence. However, we should not be fooled into believing that any reduced entropy $H_\star$ will lead to an enhancement in the probability to find self-replicators by chance: the distribution $p_\star$ needs to be close to the distribution of actual replicators. For example, omitting the instruction `x' (the {\tt h-divide} instruction that splits off a completed copy, see Table~\ref{avidainst}) certainly leads to an entropy less than one, but using such a biased distribution cannot net a self-replicator as {\tt h-divide} is required for replication.

\begin{figure}[!htbp] 
   \centering
   \includegraphics[width=4in]{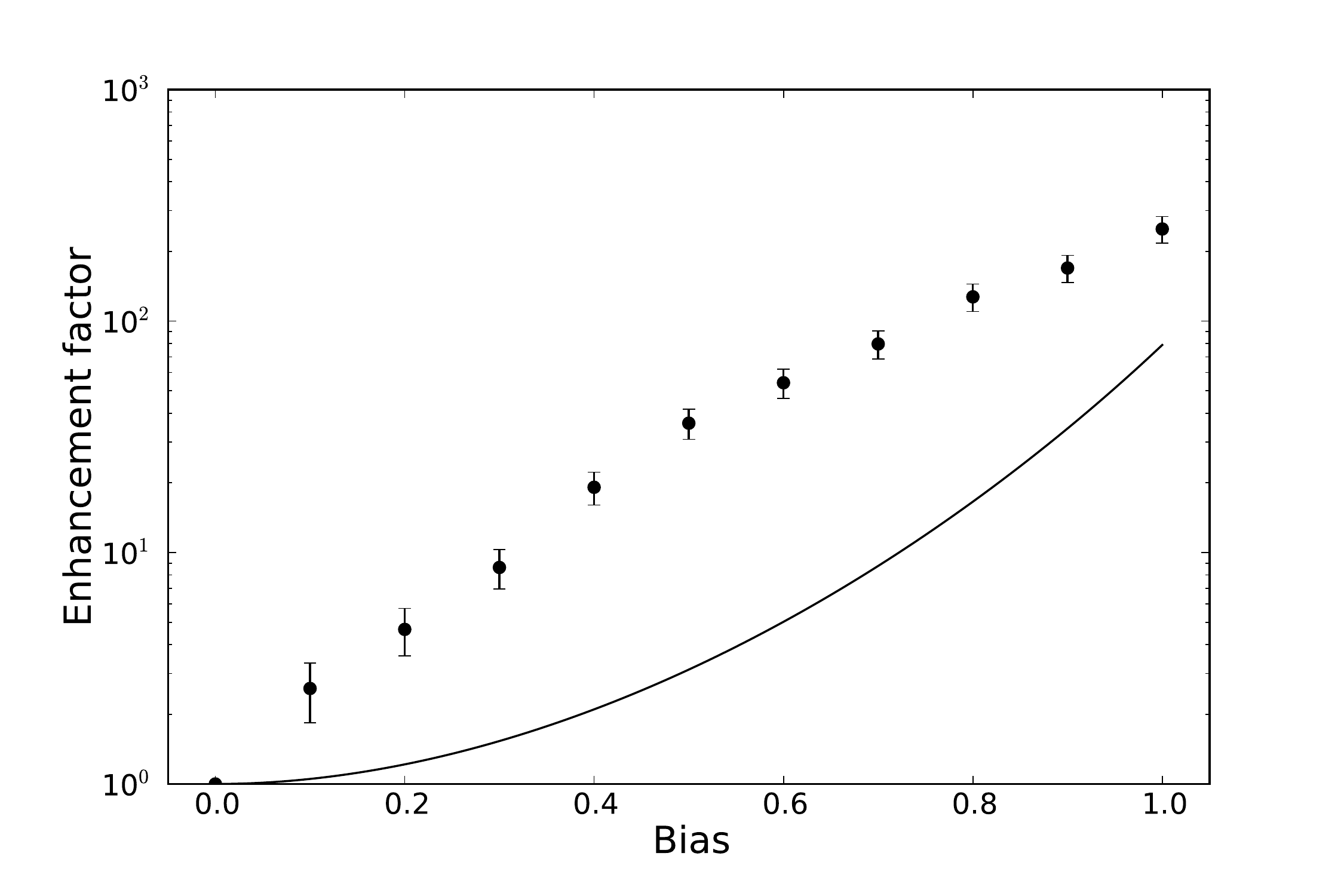} 
   \caption{The enhancement factor $E$ to find self-replicators for genomes of 15 instructions as a function of the bias, using an interpolated probability distribution $p(i,b)$. Here, $b=0$ means unbiased, and $b=1$ uses a fully biased distribution $p_\star$. Black circles represent estimates (calculated as the number of self-replicators per $10^8$ for a biased distribution divided by the number of self-replicators per $10^8$ for a uniform distribution), while error bars are standard deviations. The solid line is the naive prediction given by Eq.~(\ref{enh}).
   \label{bias15}}
\end{figure}

We proceeded to test Eq.~(\ref{enh}), by searching for self-replicators using the biased distribution $p_\star$ (see Methods). Among a billion sequences of $L=15$ generated in this manner, we found 14,495 self-replicators, an enhancement of $E=14,495/58\approx 250$, while Eq.~(\ref{enh}) predicted an enhancement of $E=81.3$. We also tested whether changing the probability distribution from uniform gradually towards $p_\star$ leads to a gradual increase in the $E$. The empirical enhancement factor shown in Fig.~\ref{bias15} indeed increases with the bias, and is larger than the one predicted from the simple approximation (\ref{enh}). This difference is likely due to a number of effects. On the one hand, we are neglecting any higher order correlations in Eq.~(\ref{full}). On the other hand, we are assuming that $H_\star\approx H(i)$ for all $i$, that is, that the entropy at each site is the same. This is not at all true for functional sequences that can be aligned (see, e.g.,~\cite{AdamiCerf2000,Adamietal2000,GuptaAdami2014}). Sequences that are obtained from a random procedure (rather than from an evolutionary process) are likely difficult to align, and therefore $H_\star\approx H(i)$ may hold. 

\begin{figure}[htbp] 
   \centering
   \includegraphics[width=\textwidth]{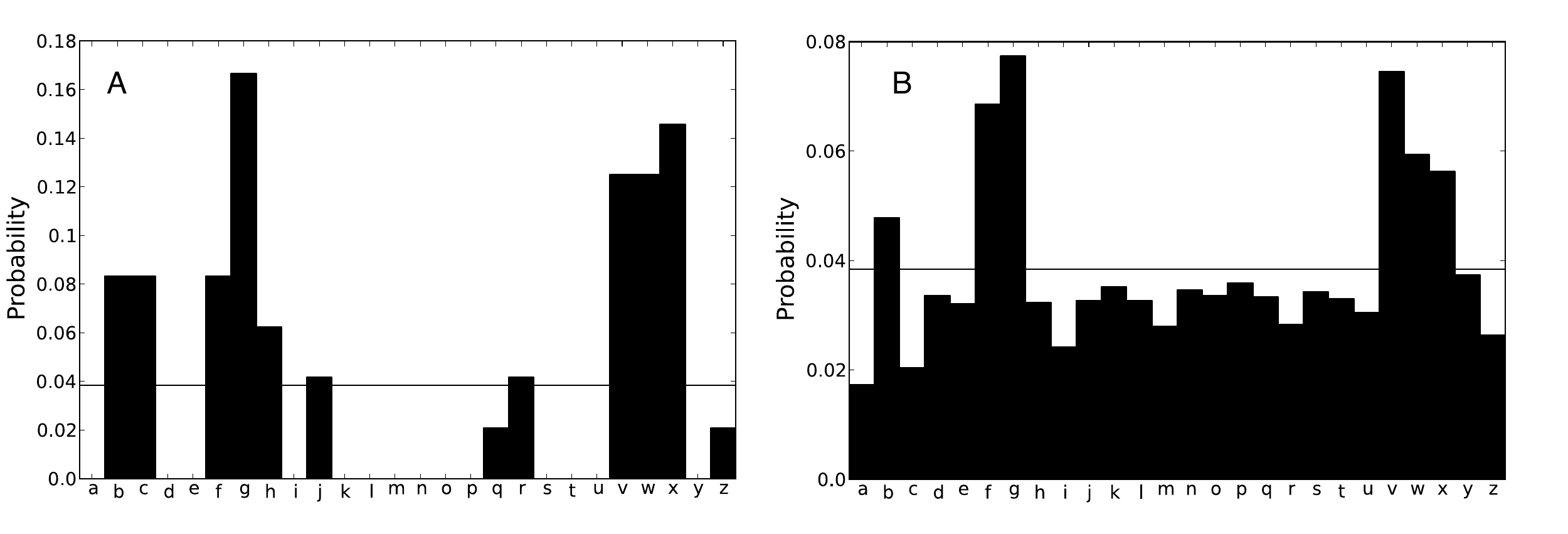} 
   \caption{Probability distribution of instructions. A: $p_\star(8)$ obtained from the replicators of length $L=8$, giving rise to an entropy $H_\star(8)=0.71$ mers. B: $p_\star(30)$ obtained from the replicators of length $L=30$, giving rise to an entropy $H_\star(30)=0.98$ mers. The solid horizontal line denotes the uniform probability distribution $1/26$ in both panels.}
   \label{prob8-30}
\end{figure}

\begin{figure}[htbp] 
   \centering
   \includegraphics[width=4in]{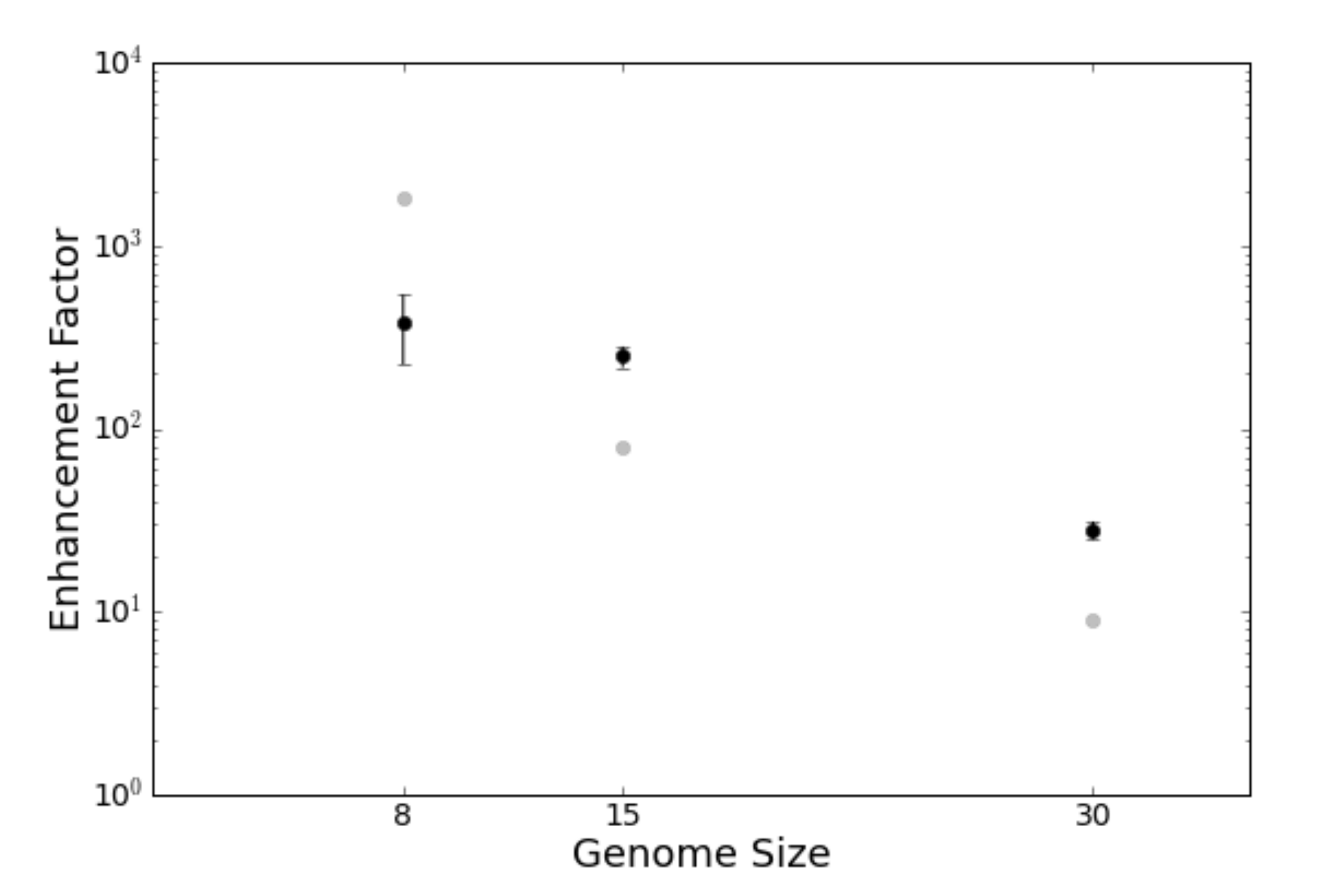} 
   \caption{Empirical enhancement factor (black dots, with 1$\sigma$ counting error), along with the predicted enhancement factor using the entropy of the distribution based on Eq.(\ref{enh}) (grey dots) for $L=8,15,30$.}
   \label{enh8-30}
\end{figure}

The enhancement works for sequences of any length, but depends on how well the biased distribution represents actual functional replicators. For example, as we found only 6 self-replicators of length 8, the distribution $p_\star(8)$ is fairly coarse (see Fig.~\ref{prob8-30}A), while the distribution we obtained from the 106 $L=30$ replicators has a significant uniform contribution (Fig.~\ref{prob8-30}B), because among the 30 instructions only a handful need to carry information in order for the sequence to be able to replicate. We show in Fig.~\ref{enh8-30} the enhancement achieved by biasing the search for each of the three length classes $L=8,15,$ and 30.

Could we use the probability distribution for sequences obtained in one length group to bias the search in another length group? Such a procedure might be useful if the statistics of monomer usage is poor (as for the case $L=8$), or if the distribution was obtained from a sequence with too much entropy (as for the case $L=30$). It turns out that this is not the case: biasing the $L=30$ search using $p_\star(15)$ does not work well (144.3 replicators found per $10^8$) compared to biasing with the ``native" $p_\star(30)$ (297 per $10^8$). In the same manner, biasing the $L=8$ search works best with the ``native" bias $p_\star(8)$, yielding 230 per $10^8$, as opposed to only 15.8 per $10^8$ biasing with $p_\star(15)$.

\begin{figure}[htbp] 
   \centering
   \includegraphics[width=\textwidth]{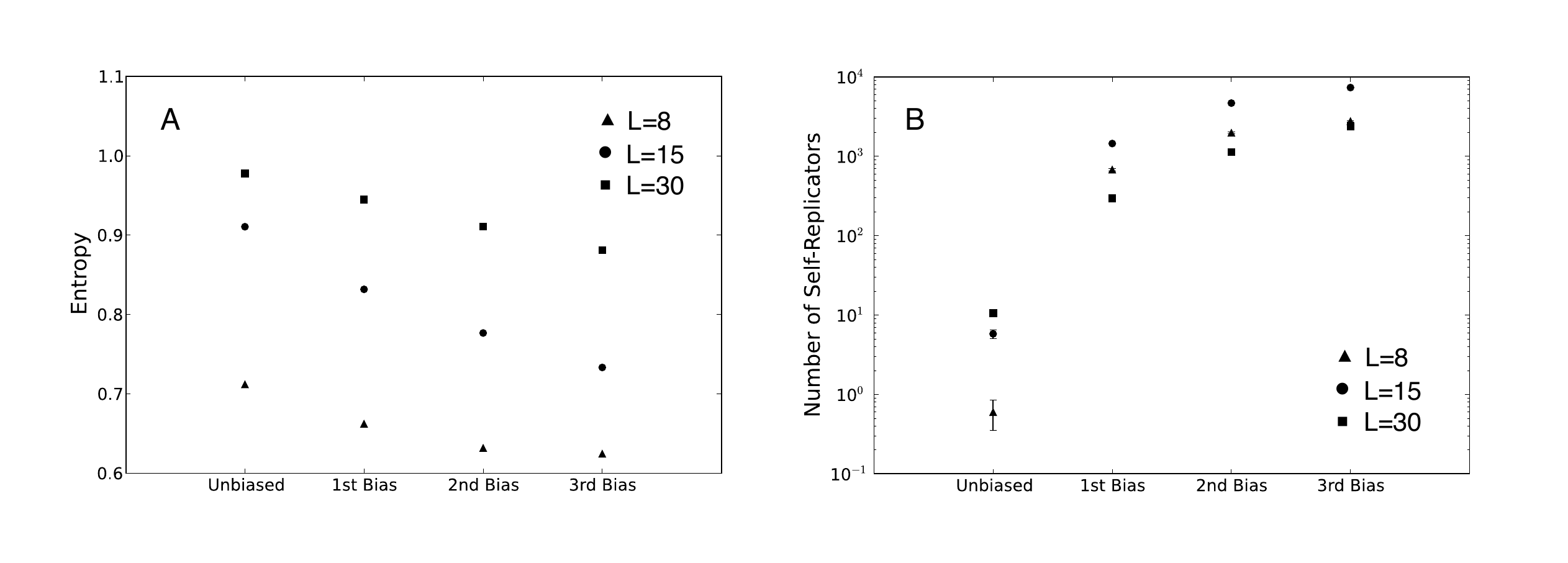} 
   \caption{A: Average per-site entropy $H_\star$ for replicators in different length classes, at various stages of biasing. ``Unbiased" reports the average per-site entropy obtained from the self-replicators that were found in an unbiased search, and whose biased distribution was used to find the self-replicators whose average per-site entropy is shown in ``1st Bias". Those in turn were used for a biased search that gave rise to replicators with bias shown in ``2nd Bias", and so on. B: Number of self-replicators (per billion) found at each biasing stage. Biasing the distribution with more ``focused" probability distributions $p_\star$ leads to an increasing yield of self-replicators, albeit with a diminishing return. In re-biasing with $L=8$, some duplicate sequences were obtained, and those are not included in the count.}
   \label{rebias}
\end{figure}

Finally we asked whether taking the self-replicators obtained from a biased search (and that consequently nets many more replicators) gives rise to a more accurate probability distribution $p_\star$, which then could be used for a more `targeted' biased search. By ``rebiasing" successively (see Methods), we did indeed obtain more and more replicators, albeit with diminishing returns (see Fig.~\ref{rebias}).

{\bf Discussion}
One of the defining characteristics of life (perhaps {\em the} defining characteristic) is that life encodes information, and information leaves a trace in the monomer abundance distribution (a non-random frequency distribution)~\cite{Dornetal2011,DornAdami2011} of the informational polymers. As life evolves, the information contained in it increases on average~\cite{Adamietal2000},  but evolution cannot explain where the first bits came from. Information can in principle arise by chance, just as an English word can appear by chance within an ASCII string that is created randomly, as per the ``dactylographic monkeys" metaphor. The ``infinite monkey theorem"  posits that a million monkeys typing on a million keyboards, if given enough time (and typing randomly) could ultimately type out all of Shakespeare's works. However, the theorem is misleading, as even correctly typing out the first 30 characters of Hamlet's soliloquy (``To be or not to be...") cannot occur during the time our universe has been around (about $4.36\times 10^{17}$ seconds), as Hamlet's 30-mer is one in about $3\times 10^{42}$. Using biased typewriters will not allow the monkeys to finish either, as it is only accelerating the search by a factor $E\approx 46,700$.

We can ask whether more sophisticated methods of biasing exist. One look at Eq.(\ref{full}) suffices to answer this question in the positive. We could begin by generating sequences biased in such a way that the more common 2-mers are generated with increased likelihood. In English text, for example, the ``dimers" `th', `he', and `in' appear with frequencies 3.56\%, 3.07\%, and 2.43\% respectively, which are significantly larger than the random dimer expectation $\approx 0.15\%$. Indeed, as the frequency of `or' is 1.28\%, while `ig' appears at 0.255\%, our 7-mer {\tt origins} would be found fairly fast. Likewise, in our 6 replicators of length $L=8$ the dimer {\tt gb} appears significantly more often than expected by the product of the likelihood of {\tt g} and {\tt b}.

Such biased search procedures can also accelerate the search for functional biomolecules where the target is a function other than self-replication. For example, when designing random peptide libraries (either for screening purposes or to perform directed evolution), researchers often bias the codons in such a way that the stop codon is rare (so-called NNB or NNS/NNK libraries~\cite{Barbasetal1992}). 
Hackel et al.~\cite{Hackeletal2010} went beyond such simple biases and constructed a protein library to screen for binding to a set of 7 targets. To bias the random sequences, they mimicked the amino acid distribution in human and mouse CDR-H3 loops (complementarity determining regions, which are found in antibodies), and found that such a library outcompetes even NNB libraries significantly: of the 20 binders that they found, 18 were traced back to the CDR-biased library.  

The implications of the present theoretical and computational analysis of the emergence of informational ``molecules" by chance for the problem of understanding the origin of life are straightforward. It is well known that monomers do not form spontaneously at the same rate. The abiotic distribution of amino acids is heavily skewed both in spark synthesis experiments as well as in meteorites~\cite{Dornetal2011}, and the same is true for other monomers such as carboxylic acids, and many other candidate alphabets in biochemistry. In many cases, the abiotic skew (often due to thermodynamic considerations) will work against the probability of spontaneous emergence of information, but in some cases it may work in its favor. In particular, we might imagine that in complex geochemical environments the abiotic distributions can be significantly different in one environment compared to another, raising the chance of abiogenesis in one environment and lowering it in another. 

We also immediately note that in chemistries where molecules do not self-replicate but catalyze the formation of other molecules, the abundance distribution of monomers would change in each catalysis step. If these monomers are recycled via reversible polymerization~\cite{Walkeretal2012}, then the activity of the molecules can change the entropy of monomers, which in turn changes the likelihood of spontaneous discovery. Should this process ``run in the right direction", it is possible that self-replicators are the inevitable outcome. This hypothesis seems testable in digital life systems such as Avida.

 {\bf Methods}
In order to explore the spontaneous emergence of self-replicators in Avida, we generated random genomes of length $L$. These genome sequences were generated with different probability distributions for the avidian instructions (we used Avida version 2.14, which can be downloaded from https://github.com/devosoft/avida). First, we generated $10^9$ random genomes for lengths $L=\{8,15,30\}$ and $3\times10^8$ sequences for $L=100$ with an unbiased (that is, uniform) instruction distribution 1/26 (because there are 26 possible instructions). In order to decide whether a genome could successfully self-replicate, we performed two tests. First, we checked whether the organism would successfully divide within its lifespan. Here, we used the traditional Avida parameters for an organism's lifespan: it must divide before it executes $20\times L$ instructions. While this indicates that an avidian could successfully reproduce, it does not imply that the avidian's descendants could also reproduce. In our search we found many viable avidians that would successfully divide into two non-viable organisms. Therefore, we only counted avidians that could self-replicate and produce offspring that could also self-replicate as true self-replicators (in other words, they are ``colony-forming"). This does not mean that every self-replicator would produce a perfect copy of itself in the absence of mutation; in fact, most of these replicators undergo implicit mutations solely due to their genome sequence, and their offspring differ in length from the parent~\cite{LaBaretal2015}. In analyzing a genome's ability to self-replicate, we used the default Avida settings,  described for example in~\cite{Ofriaetal2009}.

Next, we generated random genome sequences with a biased instruction distribution. These biased distributions were calculated by altering the probability that each instruction was generated by our random search. The probability of an instruction $i$ being generated for a biased search was set at 
\be
p(i,b)=(1-b)(1/26)+b p_\star(i),
\ee
where $b$ is the bias, $0\le b\le1$, and $p_\star(i)$ is the probability that instruction $i$ appears in the set of all genomes that were classified as self-replicators in the unbiased search. When $b=0$, the distribution is the uniform distribution and when $b=1$, the distribution is the frequency distribution for the instructions in the set of self-replicators $p_\star$ found with the unbiased search for a given length. The parameter $b$ allows us to set the bias, and thus the entropy, of the distribution to detect the role of the instruction entropy in determining the likelihood of spontaneous self-replicator emergence. For genomes of $L=15$, we generated $10^9$ random sequences with $b=1$ and $10^8$ random sequences with $b=\{0.1, 0.2, 0.3, 0.4, 0.5, 0.6, 0.7, 0.8, 0.9\}$.

Finally, we performed searches where we iteratively biased the distribution of instructions. First, we generated self-replicators with an unbiased instruction distribution. We then created another set of self-replicators with a biased distribution of instructions using the above equation with $b=1$ (referred to as ``1st bias"). However, as opposed to stopping the self-replicator generation process, we then searched for self-replicators two more times (referred to as `2nd bias' and `3rd bias'). Each time, we used the set of self-replicators from the previous bias: the distribution of instructions for the 2nd bias was derived from the set of self-replicators obtained from the 1st bias, and the distribution of instructions for the 3rd bias was derived from the set of self-replicators from the 2nd bias (in both of these we set $b=1$). We generated $10^8$ random genomes using the 1st bias for $L=\{8,30\}$ and $10^8$ random genomes using the 2nd and 3rd bias for $L=\{8,15,30\}$ with a biased instruction distribution. For $L=15$, we used the $10^9$ random genomes described above to obtain the 1st bias.  

\subsubsection*{Acknowledgements}We thank Arend Hintze and Charles Ofria for extensive discussions, and Piet Hut and Jim Cleaves for the suggestion to carry out the kind of computational work presented here. This work was supported by the National Science Foundation's BEACON Institute for the Study of Evolution in Action under contract No. DBI-0939454. We gratefully acknowledge the support of the Michigan State University High Performance Computing Center and the Institute for Cyber Enabled Research (iCER).

 \bibliography{OLEB}

\begin{thebibliography}{10}
\expandafter\ifx\csname urlstyle\endcsname\relax
  \providecommand{\doi}[1]{doi:\discretionary{}{}{}#1}\else
  \providecommand{\doi}{doi:\discretionary{}{}{}\begingroup
  \urlstyle{rm}\Url}\fi

\bibitem{Morowitz2004}
Morowitz, H.  2004 \emph{Beginnings of Cellular Life: {Metabolism Recapitulates
  Biogenesis}}.
\newblock New Haven, CT: Yale University Press.

\bibitem{Deamer1994}
Deamer, D.  1994 \emph{Origins of Life: The Central Concepts}.
\newblock Sudbury, MA: Jones \& Bartlett Publishers.

\bibitem{DeDuve1995}
deDuve, C.  1995 \emph{Vital Dust: Life as a Cosmic Imperative}.
\newblock New York, N.Y.: Basic Books.

\bibitem{Koonin2011}
Koonin, E.  2011 \emph{The Logic of Chance: The Nature and Origin of Biological
  Evolution}.
\newblock Upper Saddle River, NJ: FT Press.

\bibitem{Pateletal2015}
Patel, B.~H., Percivalle, C., Ritson, D.~J., Duffy, C.~D. \& Sutherland, J.~D.
  2015 Common origins of rna, protein and lipidprecursors in a cyanosulfidic
  protometabolism.
\newblock \emph{Nature Chemistry} \textbf{7}, 301--307.

\bibitem{Daviesetal2009}
Davies, P. C.~W., Benner, S.~A., Cleland, C.~E., Lineweaver, C.~H., McKay,
  C.~P. \& Wolfe-Simon, F.  2009 Signatures of a shadow biosphere.
\newblock \emph{Astrobiology} \textbf{9}, 241--9.
\newblock \doi{10.1089/ast.2008.0251}.

\bibitem{Vetsigianetal2006}
Vetsigian, K., Woese, C. \& Goldenfeld, N.  2006 Collective evolution and the
  genetic code.
\newblock \emph{Proc Natl Acad Sci U S A} \textbf{103}, 10,696--701.
\newblock \doi{10.1073/pnas.0603780103}.

\bibitem{Smith2008}
Smith, E.  2008 Thermodynamics of natural selection {I}: Energy flow and the
  limits on organization.
\newblock \emph{J Theor Biol} \textbf{252}, 185--97.
\newblock \doi{10.1016/j.jtbi.2008.02.010}.

\bibitem{England2013}
England, J.~L.  2013 Statistical physics of self-replication.
\newblock \emph{J Chem Phys} \textbf{139}, 121,923.
\newblock \doi{10.1063/1.4818538}.

\bibitem{Segreetal2000}
Segr\'e, D., Ben-Eli, D. \& Lancet, D.  2000 Compositional genomes: prebiotic
  information transfer in mutually catalytic noncovalent assemblies.
\newblock \emph{Proc. Natl. Acad. Sci. USA} \textbf{97}, 4112--4117.

\bibitem{NowakOtsuki2008}
Nowak, M.~A. \& Ohtsuki, H.  2008 Prevolutionary dynamics and the origin of
  evolution.
\newblock \emph{Proc Natl Acad Sci U S A} \textbf{105}, 14,924--7.
\newblock \doi{10.1073/pnas.0806714105}.

\bibitem{Vasasetal2012}
Vasas, V., Fernando, C., Santos, M., Kauffman, S. \& Szathm{\'a}ry, E.  2012
  Evolution before genes.
\newblock \emph{Biol Direct} \textbf{7}, 1.
\newblock \doi{10.1186/1745-6150-7-1}.

\bibitem{Walkeretal2012}
Walker, S.~I., Grover, M.~A. \& Hud, N.~V.  2012 Universal sequence
  replication, reversible polymerization and early functional biopolymers: a
  model for the initiation of prebiotic sequence evolution.
\newblock \emph{PLoS One} \textbf{7}, e34,166.
\newblock \doi{10.1371/journal.pone.0034166}.

\bibitem{Mathisetal2015}
Mathis, C., Bhattacharya, T. \& Walker, S.~I.  2015.
\newblock The emergence of life as a first order phase transition.
\newblock arXiv:1503.02777.

\bibitem{LazcanoMiller1996}
Lazcano, A. \& Miller, S.~L.  1996 The origin and early evolution of life:
  prebiotic chemistry, the pre-rna world, and time.
\newblock \emph{Cell} \textbf{85}, 793--8.

\bibitem{Benneretal2004}
Benner, S.~A., Ricardo, A. \& Corrigan, M.~A.  2004 Is there a common chemical
  model for life in the universe?
\newblock \emph{Curr. Opin. Chem. Biol.} \textbf{8}, 672--689.

\bibitem{Adami1998}
Adami, C.  1998 \emph{Introduction to Artificial Life}.
\newblock Berlin, Heidelberg, New York: Springer Verlag.

\bibitem{Mayfield2013}
Mayfield, J.~E.  2013 \emph{The Engine of Complexity: Evolution as
  Computation}.
\newblock New York, N.Y.: Columbia University Press.

\bibitem{Adami2002b}
Adami, C.  2002 What is complexity?
\newblock \emph{BioEssays} \textbf{24}, 1085--94.

\bibitem{Adami2012}
Adami, C.  2012 The use of information theory in evolutionary biology.
\newblock \emph{Annals NY Acad. Sci.} \textbf{1256}, 49--65.

\bibitem{Randsetal2014}
Rands, C.~M., Meader, S., Ponting, C.~P. \& Lunter, G.  2014 8.2\% of the human
  genome is constrained: variation in rates of turnover across functional
  element classes in the human lineage.
\newblock \emph{PLoS Genet} \textbf{10}, e1004,525.
\newblock \doi{10.1371/journal.pgen.1004525}.

\bibitem{Arrhenius1908}
Arrhenius, S.  1908 \emph{Worlds in the Making: The Evolution of the Universe}.
\newblock New York, NY: Harper \& Row.

\bibitem{HoyleWickramasinghe1981}
Hoyle, F. \& Wickramasinghe, N.  1981 \emph{Evolution from Space}.
\newblock New York, NY: Simon \& Schuster Inc.

\bibitem{Wickramasinghe2011}
Wickramasinghe, C.  2011 Bacterial morphologies supporting cometary panspermia:
  a reappraisal.
\newblock \emph{International Journal of Astrobiology} \textbf{10}, 25--30.

\bibitem{Pattee1961}
Pattee, H.~H.  1961 On the origin of macromolecular sequences.
\newblock \emph{Biophys J} \textbf{1}, 683--710.

\bibitem{AdamiCerf2000}
Adami, C. \& Cerf, N.~J.  2000 Physical complexity of symbolic sequences.
\newblock \emph{Physica D} \textbf{137}, 62--69.

\bibitem{Szostak2003}
Szostak, J.~W.  2003 Functional information: Molecular messages.
\newblock \emph{Nature} \textbf{423}, 689.

\bibitem{AdamiBrown1994}
Adami, C. \& Brown, C.  1994 Evolutionary learning in the {2D} {Artificial
  Life} system {Avida}.
\newblock In R.~Brooks \& P.~Maes, eds., \emph{Proceedings of the 4th
  International Conference on the Synthesis and Simulation of Living Systems
  (Artificial Life 4)}  pp. 377--381. MIT Press.

\bibitem{OfriaWilke2004}
Ofria, C. \& Wilke, C.~O.  2004 Avida: A software platform for research in
  computational evolutionary biology.
\newblock \emph{Artificial Life} \textbf{10}, 191--229.
\newblock \doi{10.1162/106454604773563612}.

\bibitem{Ofriaetal2009}
Ofria, C., Bryson, D. \& Wilke, C.  2009 Avida: A software platform for
  research in computational evolutionary biology.
\newblock In A.~Adamatzky \& M.~Komosinski, eds., \emph{Artificial Life Models
  in Software}  pp. 3--35. Springer Verlag  2nd edition.

\bibitem{Adami2006}
Adami, C.  2006 Digital genetics: unravelling the genetic basis of evolution.
\newblock \emph{Nature Reviews Genetics} \textbf{7}, 109--118.
\newblock \doi{10.1038/nrg1771}.

\bibitem{Hazenetal2007}
Hazen, R.~M., Griffin, P.~L., Carothers, J.~M. \& Szostak, J.~W.  2007
  Functional information and the emergence of biocomplexity.
\newblock \emph{Proc Natl Acad Sci U S A} \textbf{104 Suppl 1}, 8574--81.
\newblock \doi{10.1073/pnas.0701744104}.

\bibitem{Pargellis2003}
Pargellis, A.  2003 Self-organizing genetic codes and the emergence of digital
  life.
\newblock \emph{Complexity (Wiley)} \textbf{8}, 69.

\bibitem{Adami2014}
Adami, C.  2015 Information-theoretic considerations concerning the origin of
  life.
\newblock \emph{Origins of Live and Evolution of the Biospheres} \textbf{45},
  9439.

\bibitem{Lewand2000}
Lewand, R.~E.  2000 \emph{Cryptological Mathematics}.
\newblock Washington, DC: The Mathematical Association of America.

\bibitem{Reif1965}
Reif, F.  1965 \emph{Fundamentals of Statistical and Thermal Physics}.
\newblock Boston, MA: McGraw-Hill.

\bibitem{Fano1961}
Fano, R.~M.  1961 \emph{Transmission of Information: A Statistical Theory of
  Communication}.
\newblock New York and London: MIT Press and John Wiley.

\bibitem{GuptaAdami2014}
Gupta, A. \& Adami, C.  2014 Strong selection significantly increases epistatic
  interactions in the long-term evolution of a protein.
\newblock ArXiv.org preprint arXiv:1408.2761.

\bibitem{Adamietal2000}
Adami, C., Ofria, C. \& Collier, T.~C.  2000 Evolution of biological
  complexity.
\newblock \emph{Proc. Natl. Acad. Sci. U.S.A.} \textbf{97}, 4463--8.

\bibitem{Dornetal2011}
Dorn, E.~D., Nealson, K.~H. \& Adami, C.  2011 Monomer abundance distribution
  patterns as a universal biosignature: examples from terrestrial and digital
  life.
\newblock \emph{J Mol Evol} \textbf{72}, 283--95.
\newblock \doi{10.1007/s00239-011-9429-4}.

\bibitem{DornAdami2011}
Dorn, E.~D. \& Adami, C.  2011 Robust monomer-distribution biosignatures in
  evolving digital biota.
\newblock \emph{Astrobiology} \textbf{11}, 959--68.
\newblock \doi{10.1089/ast.2010.0556}.

\bibitem{Barbasetal1992}
Barbas, C.~F., 3rd, Bain, J.~D., Hoekstra, D.~M. \& Lerner, R.~A.  1992
  Semisynthetic combinatorial antibody libraries: a chemical solution to the
  diversity problem.
\newblock \emph{Proc Natl Acad Sci U S A} \textbf{89}, 4457--61.

\bibitem{Hackeletal2010}
Hackel, B.~J., Ackerman, M.~E., Howland, S.~W. \& Wittrup, K.~D.  2010
  Stability and {CDR} composition biases enrich binder functionality
  landscapes.
\newblock \emph{J Mol Biol} \textbf{401}, 84--96.
\newblock \doi{10.1016/j.jmb.2010.06.004}.

\bibitem{LaBaretal2015}
LaBar, T., Adami, C. \& Hintze, A.  2015 Does self-replication imply
  evolvability?
\newblock In P.~Andrews, L.~Caves, R.~Doursat, S.~Hickinbotham, F.~Polack,
  S.~Stepney, T.~Taylor \& J.~Timmis, eds., \emph{Proc. of European Conference
  on Artificial Life 2015}  pp. 596--602. Cambridge, MA: MIT Press.

\end{thebibliography}
 \end{document}